\documentclass[fleqn,twoside]{article}
\usepackage{multicol}
\usepackage[headings]{espcrc2}

\readRCS
$Id: espcrc2.tex,v 1.2 2004/02/24 11:22:11 spepping Exp $
\usepackage{epsfig}

\def\ph#1{{\phantom{#1}}}

\def\Poles{{\cal P}oles}

\def\d{{\rm d}}

\def\e{\epsilon}
\def\JET{J}

\title{
Infrared structure of $e^+e^- \to 3$~jets at NNLO: QED-type contributions}

\author{A.\ Gehrmann-De Ridder
\address{Institute for Theoretical Physics, ETH, CH-8093 Z\"urich,
Switzerland},
T.\ Gehrmann
\address[UZH]{Institut f\"ur Theoretische Physik,
Universit\"at Z\"urich, CH-8057 Z\"urich, Switzerland},
E.W.N.\ Glover
\address{Institute of Particle Physics Phenomenology,
        University of Durham, Durham, DH1 3LE, UK},
G.\ Heinrich
\addressmark[UZH]}

\begin{document}

\unitlength 1cm

\begin{abstract}
The NNLO QCD corrections to the $e^+e^- \to 3$~jets can be decomposed 
according to their colour factors. Out of the seven colour factors, three 
are of QED-type: $1/N^2$, $N_F/N$ and $N_F^2$. We use the antenna 
subtraction method to compute these contributions,
 providing complete expressions for the subtraction terms in 
$N_F/N$ and $N_F^2$.
\end{abstract}

\maketitle


\section{Introduction} 
Electron-positron annihilation into three jets provides one of the most
precise experimental tests of QCD dynamics. The extraction of the strong
coupling constant $\alpha_s$ from data on this 
observable~\cite{bethke} is currently limited~\cite{salam}
 by the uncertainty on 
the available theoretical next-to-leading order (NLO) calculations. 
To improve upon this situation, it is mandatory to compute 
next-to-next-to-leading order (NNLO) QCD corrections to $e^+e^-\to 3$~jets.

This calculation contains three 
ingredients~\cite{CFsquare}: matrix elements for 
$\gamma^*\to 5$~partons at tree-level,  for 
$\gamma^*\to 4$~partons at one-loop and $\gamma^*\to 3$~partons 
at two loops. The four-parton and five-parton contributions contain 
infrared singularities due to real radiation of massless final state 
partons. To extract these singularities, which cancel with explicit 
infrared singularities in the virtual corrections for any infrared-safe
observable, one needs to define a subtraction scheme at NNLO. We proposed 
a formulation of the antenna subtraction method~\cite{cullen,ant} to 
NNLO in~\cite{ourant}. 

\section{Antenna subtraction}
In the antenna subtraction method, 
antenna
functions describe the colour-ordered radiation of unresolved
partons between a
pair of hard (radiator) partons. All antenna functions at NLO and NNLO
can be derived systematically from physical matrix elements, as shown 
in~\cite{ourant,our2j}. They can be integrated over the factorized 
antenna phase space~\cite{nnlosub1} using loop integral reduction 
techniques extended to phase space integrals~\cite{am,ggh}, and then 
combined with virtual corrections to partonic processes with lower 
multiplicity.

\vspace{-0.2mm}
\section{Colour structure of $e^+e^- \to 3$~jets}
Decomposing the three-jet cross section (and related event shape observables)
at NNLO according to the QCD colour structures one finds seven colour 
factors: $N^2$, $N^0$, $1/N^2$, $N_F\,N$, $N_F/N$, $N_F^2$ and $N_{F,\gamma}$.
The last colour factor $N_{F,\gamma}$ arises from the interference of 
amplitudes with two independent quark lines coupling to the exchanged 
vector boson. This colour factor is infrared finite and numerically small
in the three-parton channel~\cite{3jme}, and contributes only a negligible 
amount to four-jet observables at NLO~\cite{4jnlo}. It can therefore be
neglected safely. 

Among the remaining six colour factors, three do not receive contributions 
from diagrams with gluon self-coupling, and are therefore of QED-type:
$1/N^2$, $N_F/N$ and $N_F^2$. In~\cite{CFsquare,ourant}, we described the 
calculation of NNLO corrections in the $1/N^2$ colour factor using 
the antenna subtraction
method. In the present paper, we describe the calculation of 
NNLO corrections in the $N_F/N$ and $N_F^2$ colour factors, thus completing 
the corrections to all QED-type colour factors. Our notation for 
partonic matrix elements, antenna functions
and subtraction terms follows the notation introduced in~\cite{ourant}.

\onecolumn
\section{Subtraction in the $N_F/N$ colour factor}
The $N_F/N$ colour factor receives contributions from five-parton 
tree-level $\gamma^* \to q \bar q q' \bar q' g$, four-parton one-loop 
$\gamma^* \to q \bar q q'\bar q'$ and $\gamma^* \to q \bar q gg$ at 
subleading colour 
as well as tree-level two-loop $\gamma^* \to q \bar q g$.
The gluon emissions are all photon-like.

\label{sec:termE}
\subsection{Five-parton contribution}
The NNLO radiation term appropriate for the three jet final state is given by
\begin{eqnarray}
{\rm d}\sigma_{NNLO,N_F/N}^R&=& 
N_{{5}}\,\frac{N_F}{N} \,  {\rm d}\Phi_{5}(p_{1},\ldots,p_{5};q) 
\,\Big[ 
B_5^{0,c}(1_q,5_g,2_{\bar q};3_{q'},4_{\bar q'})\nonumber \\
&&\hspace{0.6cm}
+B_5^{0,d}(1_q,2_{\bar q};3_{q'},5_g,4_{\bar q'})
-2\,B_5^{0,e}(1_q,2_{\bar q};3_{q'},4_{\bar q'};5_g)
\Big]
\JET_{3}^{(5)}(p_{1},\ldots,p_{5})\nonumber \\
&=& 
N_{{5}}\,\frac{N_F}{N} \,  {\rm d}\Phi_{5}(p_{1},\ldots,p_{5};q) 
\,\frac{1}{2}\,\sum_{(i,j)\in (3,4)} \Big[ 
B_5^{0,c}(1_q,5_g,2_{\bar q};i_{q'},j_{\bar q'})\nonumber \\
&&\hspace{0.6cm}
+B_5^{0,d}(1_q,2_{\bar q};i_{q'},5_g,j_{\bar q'})
-2\,B_5^{0,e}(1_q,2_{\bar q};i_{q'},j_{\bar q'};5_g)
\Big]
\JET_{3}^{(5)}(p_{1},\ldots,p_{5})\,,
\label{eq:sigrE}\end{eqnarray}
where the symmetrisation over the momenta of the secondary quark-antiquark
pair exploits the fact that the jet algorithm does not distinguish 
quarks and antiquarks. This symmetrisation reduces the number of
non-vanishing unresolved limits considerably, since the 
interference term in $B_5^{0,e}$ is odd under this interchange.
The subtraction term reads:
\begin{eqnarray}
\lefteqn{{\rm d}\sigma_{NNLO,N_F/N}^S=
N_{{5}}\,\frac{N_F}{N} \,  {\rm d}\Phi_{5}(p_{1},\ldots,p_{5};q) 
\,\frac{1}{2}\,\sum_{(i,j)\in (3,4)} \Bigg\{ } \nonumber \\
\ph{(a)} &&- A^0_3(1_q,5_g,2_{\bar q})\,
{B}^{0}_{4} (\widetilde{(15)}_q,\widetilde{(25)}_{\bar q},
i_{q'},j_{\bar q'})\,
{J}_{3}^{(4)}(\widetilde{p_{15}},\widetilde{p_{25}},p_i,p_{j}) 
\nonumber \\
\ph{(b)} &&- A^0_3(i_{q'},5_g,j_{\bar q'})\,
{B}^{0}_{4} (1_{q},2_{\bar q},\widetilde{(i5)}_{q'},\widetilde{(j5)}_{\bar q'})
\,{J}_{3}^{(4)}(p_1,p_2,\widetilde{p_{i5}},\widetilde{p_{j5}}) 
\nonumber \\
\ph{(c)} &&-  \frac{1}{2} \left\{
E^0_3(1_q,i_{q'},j_{\bar q'})\,
\tilde{A}^0_{4}(\widetilde{(1j)}_q,\widetilde{(ij)}_g,5_g,2_{\bar q})\,
{J}_{3}^{(4)}(\widetilde{p_{1j}},p_2,\widetilde{p_{ij}},p_5) 
+  (1\leftrightarrow 2)\right\}
 \nonumber \\
\ph{(d,e)} &&-  \Bigg(\, B_4^0(1_q,i_{q'},j_{\bar q'},2_{\bar q})
- \frac{1}{2} \left\{E_3^0(1_q,i_{q'},j_{\bar q'})\,
A_3^0( \widetilde{(1j)}_q,\widetilde{(ij)}_g,2_{\bar q})
+  (1\leftrightarrow 2)\right\}
\,\Bigg)
\nonumber \\
&& \hspace{1cm}
\times
{A}^0_{3}(\widetilde{(1ij)}_q,5_g,\widetilde{(2ij)}_{\bar q})
\,
{J}_{3}^{(3)}(\widetilde{p_{1ij}},\widetilde{p_{2ij}},p_5) \nonumber \\
\ph{(h)} &&+ \frac{1}{2} \Bigg\{ E_3^0(1_q,i_{q'},j_{\bar q'}) \,
A_3^0(\widetilde{(1j)_q},5_g,2_{\bar q})\, 
A^0_{3}(\widetilde{((1j)5)}_q,\widetilde{(ij)}_g,
\widetilde{(25)}_{\bar q})\, {J}_{3}^{(3)}
(\widetilde{p_{(1j)5}},\widetilde{p_{25}},\widetilde{p_{ij}}) 
+ (1\leftrightarrow 2) \Bigg\} 
\nonumber \\
\ph{(f,g)} &&- \frac{1}{2} \Bigg\{
\Bigg( \tilde{E}_4^0(1_q,i_{q'},j_{\bar q'},5_g) 
 - A_3^0(i_{q'},5_g,j_{\bar q'})
 E_3^0(1_{q},\widetilde{(i5)}_{q'},\widetilde{(j5)}_{\bar q'})\Bigg)
\nonumber \\
&& \hspace{1cm}
\times
{A}^0_{3}(\widetilde{(1ij)}_q,\widetilde{(5ij)}_g,2_{\bar q})\, 
{J}_{3}^{(3)}(\widetilde{p_{1ij}},p_2,\widetilde{p_{5ij}})
+ (1\leftrightarrow 2)
\Bigg\}  
\Bigg\}
\label{eq:nnlosubcfnf}
\end{eqnarray}

\subsection{Four-parton contribution}
The four parton contribution to the $N_F/N$ colour factor reads:
\begin{eqnarray}
{\rm d}\sigma_{NNLO,N_F/N}^{V,1}&=&
N_{{4}}\,\frac{N_F}{N}\, \left(\frac{\alpha_s}{2\pi}\right)\,
{\rm d}\Phi_{4}(p_{1},\ldots,p_{4};q)\,
\Bigg\{ - \frac{1}{2}\sum_{(i,j)\in (3,4)}
\Big( B_{4}^{1,b} (1_q,i_{q'},j_{\bar q'},2_{\bar q}) 
 \nonumber \\ &&
+ 2 C_{4}^{1,f} (1_q,i_{q},j_{\bar q},2_{\bar q}) +
\tilde A_{4}^{1,c} (1_q,i_g,j_g,2_{\bar q}) \Big)
\Bigg\}
\,
\JET_{3}^{(4)}(p_{1},\ldots,p_{4})
\,.
\label{eq:sigvE}
\end{eqnarray}
The expression is symmetrised 
over the momenta $(3)$ and $(4)$ to remove terms which are 
antisymmetric under charge conjugation, and can not be accounted for 
properly by the quark-gluon antenna functions.

The corresponding subtraction term is:
\begin{eqnarray}
\lefteqn{{\rm d}\sigma_{NNLO,N_F/N}^{VS,1}
= 
 N_{{4}}\,\frac{N_F}{N}\, \left(\frac{\alpha_s}{2\pi}\right)\,
 {\rm d}\Phi_{4}(p_{1},\ldots,p_{4};q)}\nonumber \\
\ph{(a,b)} && \times\Bigg\{
 \left[ {\cal A}_3^0(s_{12}) + {\cal A}_3^0(s_{34}) \right]\,    
B_{4}^{0} (1_q,3_{q'},4_{\bar q'},2_{\bar q})\,
\JET_{3}^{(4)}(p_{1},\ldots,p_{4}) \nonumber \\ 
\ph{(c)}&&
+ \frac{1}{4} \left[ {\cal E}_3^0(s_{13}) + 
{\cal E}_3^0(s_{14}) + {\cal E}_3^0(s_{23})+{\cal E}_3^0(s_{24}) \right]\,    
\tilde{A}_{4}^{0} (1_q,3_{g},4_{g},2_{\bar q})
\,\JET_{3}^{(4)}(p_{1},\ldots,p_{4})  \nonumber \\ 
\ph{(d)} &&-
\frac{1}{2}\,\bigg\{\bigg(
E_3^0(1_q,3_{q'},4_{\bar q'})\left[
\tilde{A}_3^1(\widetilde{(13)}_q,\widetilde{(34)}_g,2_{\bar q})
+{\cal A}_2^1(s_{1234})
A_3^0(\widetilde{(13)}_q,\widetilde{(34)}_g,2_{\bar q})\right]
\nonumber \\
\ph{(e)} &&\hspace{0mm}+ {\cal A}_3^0(s_{(\widetilde{13})2}) \,
E_3^0(1_q,3_{q'},4_{\bar q'}) 
\,{A}_3^0(\widetilde{(13)}_q,\widetilde{(34)}_g,2_{\bar q}) \nonumber \\
\ph{(f,g)} &&\hspace{0mm}+\left[\tilde{E}_3^1(1_q,3_{q'},4_{\bar q'})
+ {\cal A}_3^0(s_{34}) {E}_3^0(1_q,3_{q'},4_{\bar q'})\right] \,
A_3^0(\widetilde{(13)}_q,\widetilde{(34)}_g,2_{\bar q})\bigg)
\JET_{3}^{(3)}(\widetilde{p_{13}},\widetilde{p_{34}},p_2) 
+ (1\leftrightarrow 2)\bigg\}
\nonumber \\
\ph{(h,i)} &&-
\frac{1}{2}\,\sum_{(i,j)\in (3,4)} \bigg(
A_3^0(1_q,i_{g},2_{\bar q})
\hat A_3^1(\widetilde{(1i)}_q,j_g,\widetilde{(2i)}_{\bar q})
+\bigg[\hat A_3^1(1_q,i_{g},2_{\bar q})
\nonumber \\
\ph{(j)} 
&&\hspace{0mm}+\frac{1}{2}\left({\cal E}_3^0(s_{1i})+{\cal E}_3^0(s_{1j})
+{\cal E}_3^0(s_{2i})+{\cal E}_3^0(s_{2j})\right)
A_3^0(1_q,i_{g},2_{\bar q})
\bigg]\,
A_3^0(\widetilde{(1i)}_q,j_g,\widetilde{(2i)}_{\bar q})\nonumber \\
\ph{(k)} &&
\hspace{0mm}+ 2b_{0,F} \log \frac{q^2}{s_{12i} }A_3^0(1_q,i_{g},2_{\bar q})
A_3^0(\widetilde{(1i)}_q,j_g,\widetilde{(2i)}_{\bar q})\bigg)
\JET_{3}^{(3)}(\widetilde{p_{1i}},\widetilde{p_{2i}},p_j)
\Bigg\}
\end{eqnarray}
The terms in the second and third line of this equation, as well as in the 
second-last line are obtained by integrating terms from the five-parton 
subtraction term~(\ref{eq:nnlosubcfnf}). All remaining terms are integrated 
to yield three-parton contributions.

\subsection{Three-parton contribution}
The three parton contribution to the $N_F/N$ colour factor consists of the 
three-parton virtual two-loop correction and the integrated 
five-parton tree-level and 
four-parton one-loop subtraction terms, which read
\begin{eqnarray}
\lefteqn{{\rm d}\sigma_{NNLO,N_F/N}^{S}+{\rm d}\sigma_{NNLO,N_F/N}^{VS,1} = 
\frac{N_F}{N}} \,\nonumber \\
&& \times \Bigg\{ -  {\cal B}_4^0  (s_{12})
- \frac{1}{2}\, \tilde{{\cal E}}_4^0  (s_{13}) - 
 \frac{1}{2}\,\tilde{{\cal E}}_4^0  (s_{23}) - \frac{1}{2}
{\cal A}_3^0  (s_{12}) \, \left({\cal E}_3^0  (s_{13})
+ {\cal E}_3^0  (s_{23}) \right)
\nonumber \\ &&
- \bigg[  \hat{{\cal A}}_3^1  (s_{12})
+  \frac{1}{2}\, \tilde{{\cal E}}_3^1  (s_{13}) +  \frac{1}{2}\, 
 \tilde{{\cal E}}_3^1  (s_{23}) 
\bigg] 
\,A_3^0({1}_q,{3}_g, 2_{\bar q}) - \frac{1}{2}
 \left( {\cal E}_3^0(s_{13}) +
{\cal E}_3^0(s_{23}) \right) \, 
\tilde{A}_3^1(1_q,3_g
,2_{\bar q})\nonumber \\ 
&&
- {\cal A}_3^0(s_{12})  \, 
\hat{A}_3^1(1_q,3_g,2_{\bar q})
- \frac{b_{0,F}}{\e} \,  
{\cal A}_3^0(s_{12})
\left(({s_{12}})^{-\e} -  ({s_{123}})^{-\e}\right) \,A_3^0({1}_q,{3}_g,
2_{\bar q}) 
\Bigg\} \d \sigma_3\;.
\end{eqnarray}

Taking the infrared pole part of this expression, we obtain
cancellation of all infrared poles in this channel:
\begin{equation}
\Poles\left({\rm d}\sigma_{NNLO,N_F/N}^{S}\right) +
\Poles\left({\rm d}\sigma_{NNLO,N_F/N}^{VS,1}\right) +
\Poles\left({\rm d}\sigma_{NNLO,N_F/N}^{V,2}\right) = 0 \,.
\end{equation}

\section{Subtraction in the $N_F^2$ colour factor}
\label{sec:termF}
The $N_F^2$ colour factor receives contributions only from the four-parton 
one-loop process $\gamma^*\to q\bar q q'\bar q'$ and from the 
three-parton two-loop process  $\gamma^*\to q\bar q g$. 

\subsection{Four-parton contribution}
The four-parton one-loop contribution to this colour factor is 
\begin{eqnarray}
{\rm d}\sigma_{NNLO,N_F^2}^{V,1}&=&
N_{{4}}\,N_F^2\, \left(\frac{\alpha_s}{2\pi}\right)\,
{\rm d}\Phi_{4}(p_{1},\ldots,p_{4};q)\,
B_{4}^{1,c} (1_q,3_{q'},4_{\bar q'},2_{\bar q})\; 
\JET_{3}^{(4)}(p_{1},\ldots,p_{4}).\hspace{4mm}
\label{eq:sigvNF2}
\end{eqnarray}
This contribution is free of explicit infrared poles (as can be inferred 
from the absence of a five-parton contribution to this colour structure). 

The subtraction term appropriate to this contribution is 
\begin{eqnarray}
\label{eq:VS1NF2}
{\rm d}\sigma_{NNLO,N_F^2}^{VS,1}
&= &  N_{{4}}\, N_F^2\, \left(\frac{\alpha_s}{2\pi}\right)\,
{\rm d}\Phi_{4}(p_{1},\ldots,p_{4};q)\,
\frac{1}{2}\,\Bigg\{ \nonumber \\
&& 
\bigg[\left(
\hat E_3^1(1_q,3_{q'},4_{\bar q'})
+ 2 \,  b_{0,F} \, \log\frac{q^2}{s_{134}} \,
E_3^0(1_q,3_{q'},4_{\bar q'})\right)
\,
A_3^0(\widetilde{(13)}_q,\widetilde{(34)}_g,
2_{\bar q}) 
\nonumber \\ && \hspace{3mm} + E_3^0(1_q,3_{q'},4_{\bar q'})\,
\hat{A}_3^1(\widetilde{(13)}_q,\widetilde{(34)}_g
,2_{\bar q})
\bigg] \JET_{3}^{(3)}(\widetilde{p_{13}},\widetilde{p_{34}},p_2)
\nonumber \\ &+&
\bigg[ \left(\hat E_3^1(2_{\bar q},3_{q'},4_{\bar q'})
+ 2\, b_{0,F}\, \log\frac{q^2}{s_{234}}\,
E_3^0(2_{\bar q},3_{q'},4_{\bar q'})\right)
\,A_3^0(1_q,\widetilde{(34)}_g,
\widetilde{(23)}_{\bar q}) \nonumber \\ &&
\hspace{3mm} + E_3^0(2_{\bar q},3_{q'},4_{\bar q'})\,
\hat{A}_3^1(1_q,\widetilde{(34)}_g,\widetilde{(23)}_{\bar q})
 \bigg]
\JET_{3}^{(3)}(p_1,\widetilde{p_{34}},\widetilde{p_{23}})\Bigg\}\;.
\end{eqnarray}
Although $\hat E_3^1$ and $\hat{A}_3^1$ contain explicit infrared poles, 
these cancel in their sum, as can be seen from (5.16) and (6.32) 
of~\cite{ourant}. ${\rm d}\sigma_{NNLO,N_F^2}^{VS,1}$ is therefore free of 
explicit infrared poles.

\subsection{Three-parton contribution}

The three parton contribution to the $N_F^2$ colour factor consists of the 
three-parton virtual two-loop correction and the integrated 
four-parton one-loop subtraction term, which reads
\begin{eqnarray}
\lefteqn{{\rm d}\sigma_{NNLO,N_F^2}^{VS,1} = 
N_F^2\, 
\frac{1}{2}} \nonumber \\
&\times& 
\bigg[\left(
\hat{\cal E}_3^1(s_{13}) + \hat{\cal E}_3^1(s_{23})\right)
\,A_3^0({1}_q,{3}_g, 2_{\bar q}) + 
 \left( {\cal E}_3^0(s_{13}) +
{\cal E}_3^0(s_{23}) \right) \, 
\hat{A}_3^1(1_q,3_g
,2_{\bar q})\nonumber \\ 
&&
+ \frac{b_{0,F}}{\e} \, \left[ 
{\cal E}_3^0(s_{13})
\left(({s_{13}})^{-\e} -  ({s_{123}})^{-\e}\right) +
{\cal E}_3^0(s_{23})
\left(({s_{23}})^{-\e} -  ({s_{123}})^{-\e}\right) \right] A_3^0({1}_q,{3}_g,
2_{\bar q}) 
\bigg] \d \sigma_3\;.
\end{eqnarray}

Taking the infrared pole part of this expression, we obtain
cancellation of all infrared poles in this channel:
\begin{equation}
\Poles\left({\rm d}\sigma_{NNLO,N_F^2}^{VS,1}\right) +
\Poles\left({\rm d}\sigma_{NNLO,N_F^2}^{V,2}\right) = 0 \,.
\end{equation}

\begin{multicols}{2}

\section{Numerical implementation}

Starting from the program {\tt EERAD2}~\cite{cullen}, which computes 
the four-jet
production at NLO, we implemented the NNLO antenna subtraction method 
for all QED-type colour factor contributions to $e^+e^-\to 3j$. {\tt EERAD2}
already  contains the five-parton and four-parton 
matrix elements relevant here, as well as the NLO-type subtraction terms. 

The implementation contains three channels, classified 
by their partonic multiplicity: 
(a) in the five-parton channel, we
integrate ${\rm d}\sigma_{NNLO}^{R} - {\rm d}\sigma_{NNLO}^{S}$;
(b) in the four-parton channel, we integrate
${\rm d}\sigma_{NNLO}^{V,1} - {\rm d}\sigma_{NNLO}^{VS,1}$;
(c) in the three-parton channel, we integrate
${\rm d}\sigma_{NNLO}^{V,2} +{\rm d}\sigma_{NNLO}^{S}
+ {\rm d}\sigma_{NNLO}^{VS,1}$.
The numerical integration over these channels is carried out by Monte Carlo 
methods. 

By construction, the integrands in the four-parton and 
three-parton channel are free of explicit infrared poles. In the 
five-parton and four-parton channel, we tested the proper implementation of 
the subtraction by various checks. To verify the local convergence of 
the subtraction terms and to verify that cancellations among individual 
contributions to the subtraction terms take place as expected,
we examined the behaviour of matrix element and 
subtraction terms along various trajectories approaching the unresolved 
limits. Moreover, we checked the correctness of the 
subtraction by introducing a 
lower cut (slicing parameter) on the phase space variables, and observing 
that our results are independent of this cut (provided it is 
chosen small enough). This behaviour indicates that the 
subtraction terms ensure that the contribution of potentially singular 
regions of the final state phase space does not contribute to the numerical 
integrals, but is accounted for analytically.
Finally, distributions in double and triple invariants of the  five-parton 
or four-parton phase space illustrate the proper onset of the 
subtraction terms towards the single and double unresolved edges of phase 
space.

\section{Outlook} 
In this paper, we described the calculation of NNLO corrections 
to $e^+e^-\to 3$~jets in all QED-type colour factors: $1/N^2$, $N_F/N$ and 
$N_F^2$, using the antenna subtraction method~\cite{ourant}. 
We documented explicit expressions for the subtraction terms 
for $N_F/N$ and $N_F^2$ ($1/N^2$ was documented in~\cite{ourant} already) and
described their implementation in a numerical program computing three-jet 
cross sections and related event shape observables. Work on the 
remaining colour factors is ongoing.

\section*{Acknowledgements}
This research was supported in part by the Swiss National Science Foundation
(SNF) under contracts PMPD2-106101, 200021-101874  and 200020-109162 and
 by the UK Particle Physics and Astronomy  Research Council.

\end{multicols}

\end{document}